# A stochastic chemical dynamic approach to correlate autoimmunity and optimal vitamin-D range

**Susmita Roy, Krishna Shrinivas and Biman Bagchi***

**SSCU, Indian Institute of Science, Bangalore 560012, India.**

**(Email: profbiman@gmail.com)**

*Abstract*

**Motivated by several recent experimental observations that vitamin-D could interact with antigen presenting cells (APCs) and T-lymphocyte cells (T-cells) to promote and to regulate different stages of immune response, we developed a coarse grained kinetic model in an attempt to quantify the role of vitamin-D in immunomodulatory responses. Our kinetic model, developed using the ideas of chemical network theory, leads to a system of nine coupled equations that we solve both by direct and by stochastic (Gillespie) methods. Both the analyses consistently provide detail information on the dependence of immune response to the variation of critical rate parameters. We find that although vitamin-D plays a negligible role in the initial immune response, it exerts a profound influence in the long term, especially in helping the system to achieve a new, stable steady state. The study explores the role of vitamin-D in preserving an observed bistability in the phase diagram (spanned by system parameters) of immune regulation, thus allowing the response to tolerate a wide range of pathogenic stimulation which could help in resisting autoimmune diseases. We also study how vitamin-D affects the time dependent population of dendritic cells that connect between innate and adaptive immune responses. Variations in dose dependent response in anti-inflammatory and pro-inflammatory T-cell populations to vitamin-D correlate well with recent experimental results. Our kinetic model allows for an estimation of the range of optimum level of vitamin-D required for smooth functioning of the immune system and for control of both hyper-regulation and inflammation. Most importantly, the present study reveals that an overdose or toxic level of vitamin-D or any steroid analogue could give rise to too large a tolerant response, leading to an inefficacy in adaptive immune function.**

# Introduction

It is now certain that vitamin-D is involved in large number distinct immune responses, although our quantitative understanding of these processes at cellular level still remains largely incomplete. This is because of the enormous complexity of human immune system which depends on a large number of interacting (some may be still unknown) components. Furthermore, the immune system is not only time dependent but also individual case specific. Spurred by modern epidemiologic studies, efforts in the last two decades have been directed towards understanding the origin of non-classical immunomodulatory responses believed to be triggered by active 1, 25-dihydroxy vitamin-D [1-6]. Beyond its established classical function in calcium metabolism, studies on vitamin-D are now progressively focused on its pleiotropic actions [1-6].

Vitamin-D mediated infection remedial therapies have been followed over past 150 years. Since early 1900s, usage of cod-liver oil and UV light became widely recognized as the essential sources of vitamin-D. Therapeutic use of vitamin-D first drew attention in 1849, when William used cord- liver oil to cure over 400 tuberculosis (TB) patients [7]. After a long 50 years gap, Niels Finsen won the Nobel prize by highlighting the medicinal approach of UV exposure by which he treated over 800 patients affected by lupus vulgaris (a cutaneous form of TB) [8, 9]. In Indian traditional Auyrvedic treatments, use of sunlight to treat and reduce diseases goes back to several thousand years where it is referred to as “Suryavigyan” (Meaning: science of Sun light).

Vitamin-D plays distinct roles both in innate and adaptive immunity. Several experimental and clinical studies have revealed that endogenously produced active vitamin-D (1, 25$(OH)_2D_3$) in macrophages enhances the production rate of anti-microbial peptides (cathelicidin, β-defensins, etc), to promote the innate immunity [10, 11]. Subsequently, the conversion of 25-$D_3$ into functional 1, 25-$D_3$ (known as active vitamin-D and referred to as D*) in antigen presenting cells (APCs, such as dendritic cells, macrophages) exerts potent effect on adaptive immune system [12]. Past epidemiologic data highlight the link between vitamin-D insufficiency and a range of immune-mediated disorders namely various types of autoimmune diseases. Experimental studies on the immunomodulatory properties of vitamin-D show that autoimmunity is primarily driven by enhanced number of T helper cells (e.g. Th1) that attack various self-tissues in the body. In particular, the inhibitory effect of vitamin-D on such pro-inflammatory T-cell responses and promoting regulatory T-cells ($T_{Reg}$), may, at least in part, have begun to explain some of these associations [11-15].

Some recent experimental studies shed light on such regulatory actions exerted by both vitamin-D and regulatory T-cells and their interplay in resisting autoimmunity. However, the distinct functionalities of the effector T-cells (briefly defined in the supplementary material: S1) [16] often found to evolve in presence of antigen arrested by antigen presenting cells (APCs) that impel the appropriate co-stimulatory signals to induce mature native T-cells [17, 18]. In the year of 2000, Jonuleit and coworkers characterized different types of T cell responses that are crucially dependent on maturation phase of the dendritic cells (DCs). They reported that while stimulation by mature DCs results in a strong expansion of inflammatory Th1 cells, contacts of the native T-cells with immature DCs promote the development of IL-10, producing T cell regulatory 1-like cells [19]. In 2003, Powrie and Maloy proposed an interaction scheme by surveying such inter relation between APCs and T-cell responses [20]. During the same period of time, Piemonti and coworker mentioned about the distinct role of $1,25(OH)_2D_3$ in modulating the immune responses through the inhibition of DC differentiation and maturation into potent APCs [21]. The role of vitamin-D has also emerged into the fact that while the active form of vitamin-D adversely affects T cell activation, proliferation and differentiation, it additionally facilitates the production of regulatory T cells ($T_{Reg}$) that functions as an effective immune controller [22-24]. Research on the role of vitamin-D greatly benefitted from the recent study by Correale et al. Their experiment showed that effector T cells are able to metabolize $25(OH)D_3$ into biologically active $1,25(OH)_2D_3$ in DCs, since these T cells express 1$\alpha$-hydroxylase enzyme that constitutively facilitates this conversion [25].

In the present article we develop a theoretical coarse grained kinetic network model based on the above mentioned experimental observations. Our main objective is to explore quantitatively the dependence of immunity on vitamin-D and investigate its possible role in reducing the risk of auto-immune diseases and cancer. We analyze several immunemodulatory issues that are controlled by vitamin-D, including both innate and adaptive responses as articulated in several experimental reports and reviews [11-12, 26]. Although there are numerous other complex biochemical reactions involved in the immonological pathways, we have considered only a certain number of them that are the essential participants of immune system and have direct interaction with vitamin-D.

We also address the concern for optimal range of vitamin-D intake that has been raised by World Health Organization's international agency for research on cancer. Present study especially suggests that inhibitory action exerted by regulatory T-cell induced by vitamin-D and by vitamin-D itself on adaptive response could play an important role in prevention of autoimmune diseases. There do exist several early mathematical models that studied inflammatory roles of effector T cells and their regulation by regulatory T-cells. In recent years Friedman et al. studied the effect of T cells on inflammatory Bowel Disease [27]. Pillis and coworkers investigated effects of regulatory T-cell on renal cell carcinoma treatment [28]. In another model Villoslada et al. observed the dynamic cross regulation of antigen-specific effector and regulatory T cell subpopulations in connection with microglia in brain autoimmunity [29]. Perhaps, the most relevant model for the regulation of T-cell in the immune system was presented by a recent paper by Fouchet et al [30]. They identified the important ingredients of the immune system and formulated coupled rate equations for the entire process to show the regulation of effector and regulatory T-cells by changing various rate constants.

While all these models are neat, they did not include the essential effects of vitamin-D [27-30]. Several experiments have already shown the importance of vitamin-D in both the innate and adaptive immune system. Here we have implanted the nonlinear effect of vitamin-D in basic model of T-cell regulation. The nonlinear effect of vitamin-D comes indirectly through the pathogen after effector T-cell production. The production of effector T-cell signals again initializes the activation of vitamin D. *This model primarily seeks to understand adaptive response activation and effect of vitamin-D on the tolerance/of regulatory nature of the response*. Hence we have emphasized the regulatory function of vitamin-D in the adaptive immune system. We have assumed that the innate mechanism annihilates pathogens at constant rate by production of antimicrobial peptides and this leads to the 1$^{st}$ defense against infectious diseases.

The important constituents of the model considered here are the following: (i) pathogen/antigen/epitope/self-Antigen, (ii) naive T-cell, (iii) myeloid dendritic dell in the form of professional antigen presenting cells (APC), both in their resting and activated forms (iv) effector and regulatory T cells, (v) inactive vitamin-D ($25(OH)D_3$) and active vitamin-D ($1,25(OH)_2D_3$). However the participants, such as vitamin-D receptor (VDR) and the enzyme 25(OH)D3-1α-hydroxylase (CYP27B1) that simultaneously function over inactive vitamin-D to form activate vitamin-D ($1,25(OH)_2$ D)-VDR protein complex, are considered as implicit factors for activation of the required transcriptional motif.

It is important to emphasize here that we have essentially combined the three important experimental observations those include the essential features of the adaptive responses reported by (i) Powri and Maloy, [20] (ii) Jorge Correale et al. [25] and (iii) Lorenzo Piemonti et al. [21]. In **Figure 1** we have presented the complex interaction network model that includes various components and their inter-relation and regulation involved in the immune system.

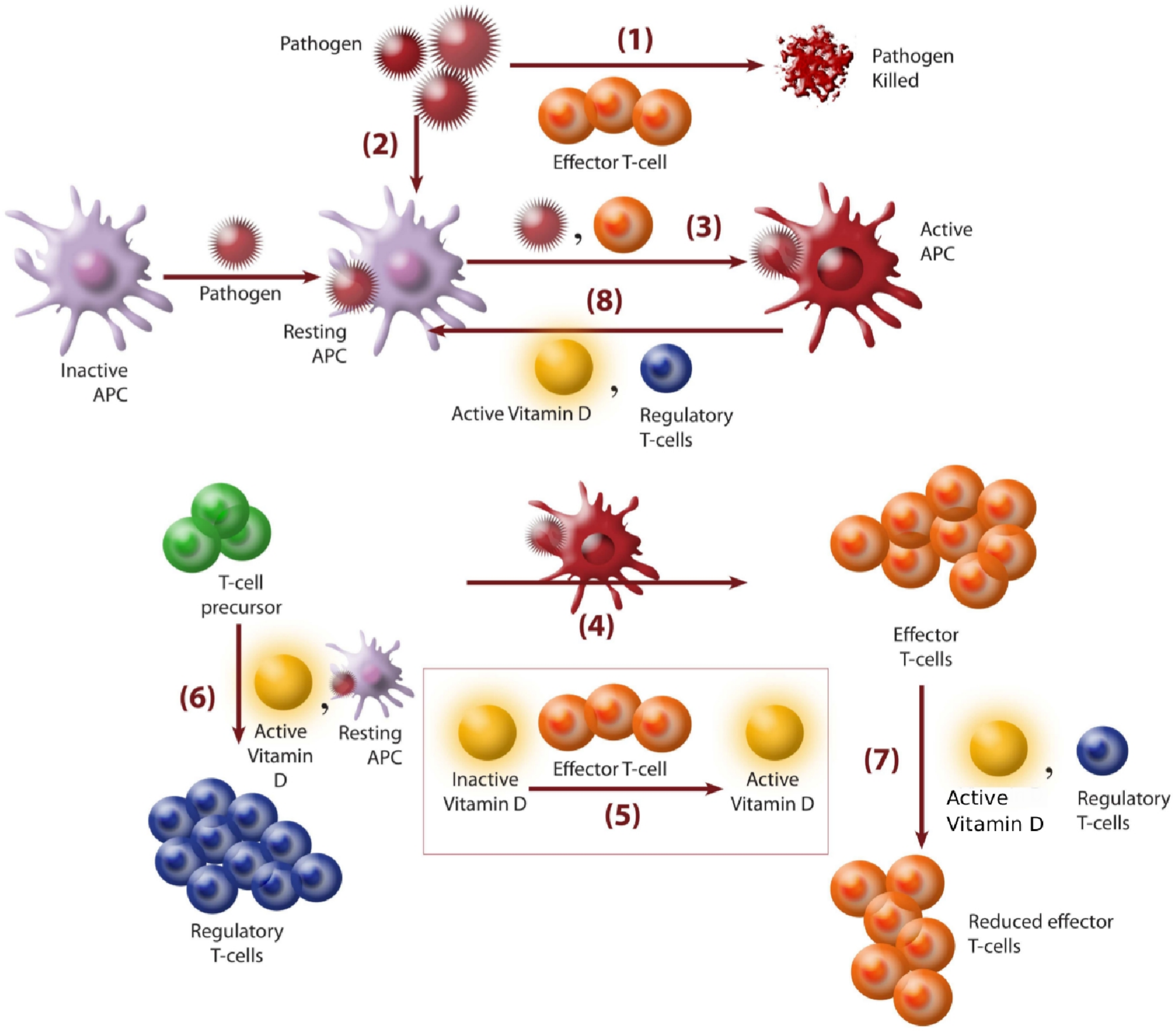

**Figure 1: A schematic representation of adaptive immune responses in terms of cellular interactions including vitamin-D, based on some experimental results and clinical observations.**

**In the scheme, the primary events are the following: (1) The main step is the annihilation of pathogen by effector T-cells [$T_{Eff}$]. (2) In presence of pathogen inactive APC becomes stimulated after pathogen recognition and form resting APC. (3) Resting APC is activated either by pathogen or by the presence of any effector T-cell [19, 20]. (4) Activation of effector T-cells are initiated by these active APC. (5) Effector T-cells initiates the formation of active vitamin-D from its inactive form [24]. (6) Resting APC and active vitamin-D both can stimulate the production of regulatory T-cell from its precursor [25]. (7) Enhancement of rate of production of effector T-cells is controlled**

**by both regulatory T-cells ($T_{Reg}$) and active vitamin-D [24]. (8) In addition, vitamin-D and regulatory T-cell up-regulate the formation of more resting APC from active APC [21].**

The present approach of network kinetic model building bears strong resemblance to similar methods adopted earlier in the study of kinetic proof reading [31, 32] and also in enzyme kinetics [33-35]. In all these studies, precise quantitative prediction is hindered by imperfect knowledge about the system parameters; especially values of rate constants are often not available. This lacuna is indeed a source of serious problem not only in study of kinetic proof reading, enzyme kinetics but also, as we find here, in theoretical investigations of immunology. Finally, master equations involved in all such problems are solved by employing the method of mean first passage time [32, 36, 37], Gillespie algorithm or straight forward numerical integration. We adopt both the deterministic approach by solving differential equations numerically and stochastic simulation by employing Gillespie algorithm [38, 39]. The coarse-graining of interaction network, elaboration of the reaction scheme and the master equations are discussed in the method section.

The values of parameters involved (rate constants and concentrations) may span a wide range, and can vary from case to case. Thus, a study of the response to the variation of the important parameters has been carried out. Such a study is clearly necessary in the present context.

## Results

Under any sort of pathogenic attack, a healthy immune system always aims at killing the foreign antigen by enhancing the proliferation and differentiation rate of effector T-cells. However, excess number of certain effector T-cells has the negative effect in such a way that these cells may fail to distinguish between body's self and foreign peptides and may start destroying self tissues. Such events are often characterized as a *weak regulation* of our immune system. Thus a healthy immune system usually functions within a balanced regulation where the concentration of effector T-cells remains within a safe range so that it can effectively overcome the pathogenic stimulation and also can resist the exploding growth of effector T-cells. The production of effector T-cell again depends on the APC activation process controlled by the two rate parameters: Rate of APC activation by pathogenic stimulation ($k_{inp}$) and rate of APC reactivation by effector T cells ($k_{rese}$). Here comes the role of vitamin-D whose optimum level effectively maintains a balanced immune regulation. Vitamin-D efficiently promotes regulatory T-cells activity. Moreover, vitamin-D itself is assigned to reduce the hyper activity of APCs and effector T-cells.

In several other model studies only regulatory T-cells are assumed to maintain such regulation [28-30]. There are several experimental and clinical observations revealing the important role of vitamin-D and its concentration dependent effects in immune regulation. However we are not aware of a single theoretical model study that has been employed to investigate such interesting role of vitamin-D.

**Effect of vitamin-D on T-cells population: Transition from weak to strong regulation**

The opposing role of regulatory and effector T-cells in immunological activity, and their respective regulation by vitamin-D often determine the strength of immune-regulation and the ultimate fate of a disease. The regulation is determined by the activation of APCs followed by the activation of effector T-cells. In the present study we have categorized the regulation into three groups based on APC and effector T-cell interaction parameter ($k_{rese}$): (i) Strong regulaion, (ii) moderate regulation and (iii) weak regulation. To investigate several vitamin-D associated factors we have performed time evolution analysis of each participating element after the pathogen attack to study their long time behavior. We have studied all these three regulation limits by varying $k_{rese}$ both in the absence and in the presence of vitamin-D at different pathogenic/antigenic stimulation ($k_{inp}$).

Numerical results from solution of our system of equations are shown in **Figure 2** as a series of curves for all the three regulation limits, both in the presence and absence of vitamin-D. The results are quite revealing and we discuss them in more detail below.

Here we find from **Figure 2(a)** that in absence of vitamin-D when system falls under a strong regulation ($k_{rese}$=10) limit, the presence of standard level of vitamin-D, in comparison, is found to preserve that strong regulation efficiently (see **Figure 2(b)**). At moderate regulation regime, vitamin-D is found to maintain the bistable region ($k_{rese}=10^2$) as well that appears in absence of vitamin-D (see **Figure 2(c)**). In an early study, Fouchet and coworker [30] analyzed the steady state values of T-cells in these three regulation regimes and showed the same interesting phenomena but in absence of vitamin-D. We have also observed such bistable region both in absence and presence of vitamin-D as shown in **Figure 2(c) and 2(d)** where both strong and weak regulation can coexist. When we shift the moderate regulation regime towards weakly regulated state ($k_{rese}=10^3$) in absence of vitamin-D (see **Figure 2(e))**, standard level of vitamin-D is still found to create a moderate regulation over the steady state population of effector T-cell (see **Figure 2(f)**). We observe that at very high $k_{rese}$ values or a very high antigenic stimulation ($k_{inp}$) the system is always found to fall in a weakly regulated regime where effector T-cells are abundant, even if there presents a standard level of vitamin-D. However vitamin-D assists to preserve the required (moderate/bistable) regulation over a long range of $k_{rese}$ and indeed exerts a control over a wide range of pathogenic strength ($k_{inp}$). Depending on the intensity of antigenic stimulation, system mounts on an effective regulation to control the inflammation. This result inevitably suggests the important role of vitamin-D in switching on such required regulation.

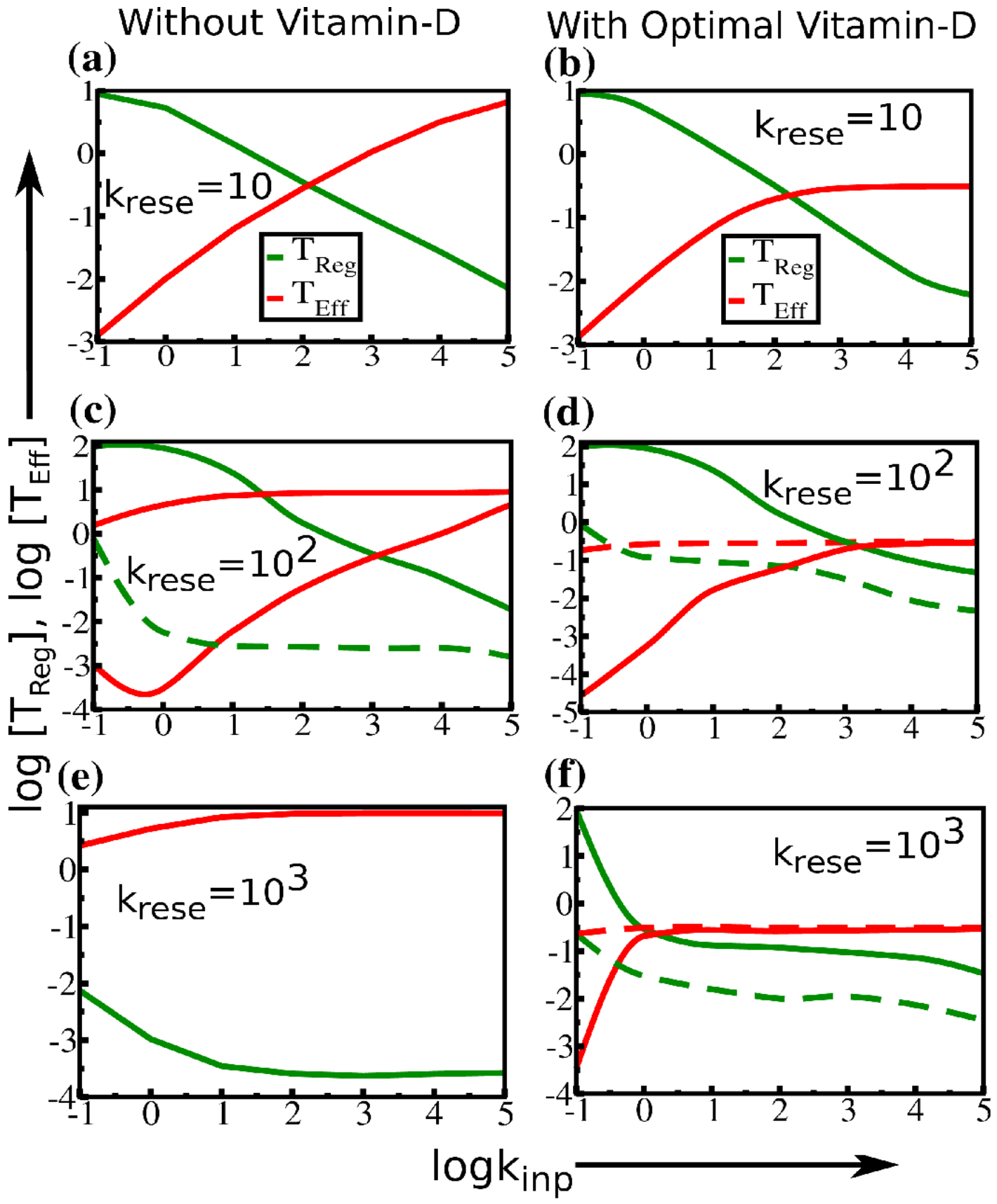

**Figure 2: Variation in T-cell concentration under weak to strong regulation.**

**Steady state concentrations of effector T-cells ($T_{Eff}$, shown in red) and regulatory T-cells ($T_{Reg}$, shown in green) are plotted against various ranges of pathogenic stimulation $(k_{inp})$ at the three different APC mediated effector T-cell regulations $(k_{rese})$. We find a stable strongly regulated state $(k_{rese} = 10)$ both (a) in absence of vitamin-D and (b) in presence of vitamin-D. The strong regulation remains strong in presence of vitamin-D. An example of bi-stable state: $(k_{rese} = 10^2)$ where both weakly regulated state (shown in dashed line) and strongly regulated state (shown in solid line) can coexist (c) in absence of vitamin-D and (d) in presence of vitamin-D. A stable weakly regulated state appears $(k_{rese} = 10^3)$ (e) in absence of vitamin-D. (f) In presence of vitamin-D, it turns out to be a bi-stable state up to certain pathogenic stimulation ($k_{inp}$) limit. Beyond that limit it falls in a weakly regulated regime. In the bi-stable situation there always exists an unstable state. Note that here we consider the vitamin-D related rate constants as, $k_{aD*} = 10^{-7}, k_{eD*} = 10^{-3}$ and the other rate values given in Table 1. Optimal vitamin-D concentration signifies the steady state value of vitamin-D (~50nmol/lit).**

Subsequent to the previous results it is worth mentioning here that bistability is a key concept for understanding the basic phenomena of cellular functioning [40, 41]. Interestingly, in presence of vitamin-D bistability is modified to be more robust and to tolerate significant changes of pathogenic stimulation. With the classification of three regulation regions (weak, moderate, strong) we investigated the boundaries in between any two. As in previous plot now we simultaneously vary both the rate of pathogenic stimulation $(k_{inp})$ at the three different APC mediated effector T-cell regulation $(k_{rese})$ (see **Figure 3**). It is necessary to point out that here the deceased production of active APCs is playing a central role in restraining the area of bistable region. In absence of vitamin-D, the inhibitory action of regulatory T-cells on active APC and thus the enhanced production of resting APCs are particularly responsible for emergence of such bistable region. In presence of vitamin-D, such degradation in active APC population enhances due to the combined effects of upregulated regulatory T-cells and Vitamin-D. **Figure 3(b)** provides a clear description that in presence of standard (optimum) level of vitamin-D, bistable region is expanded towards a larger rate of APC mediated effector T-cell activation. However, it is evident from the figure that in presence of vitamin-D a weak regulation is afar to achieve. In this case the weakly regulation regime is located towards the larger rate values of effector T-cell activation. The result signifies the strength of vitamin-D which prevents the immune system from the over-explosive limit of effector T-cell activity.

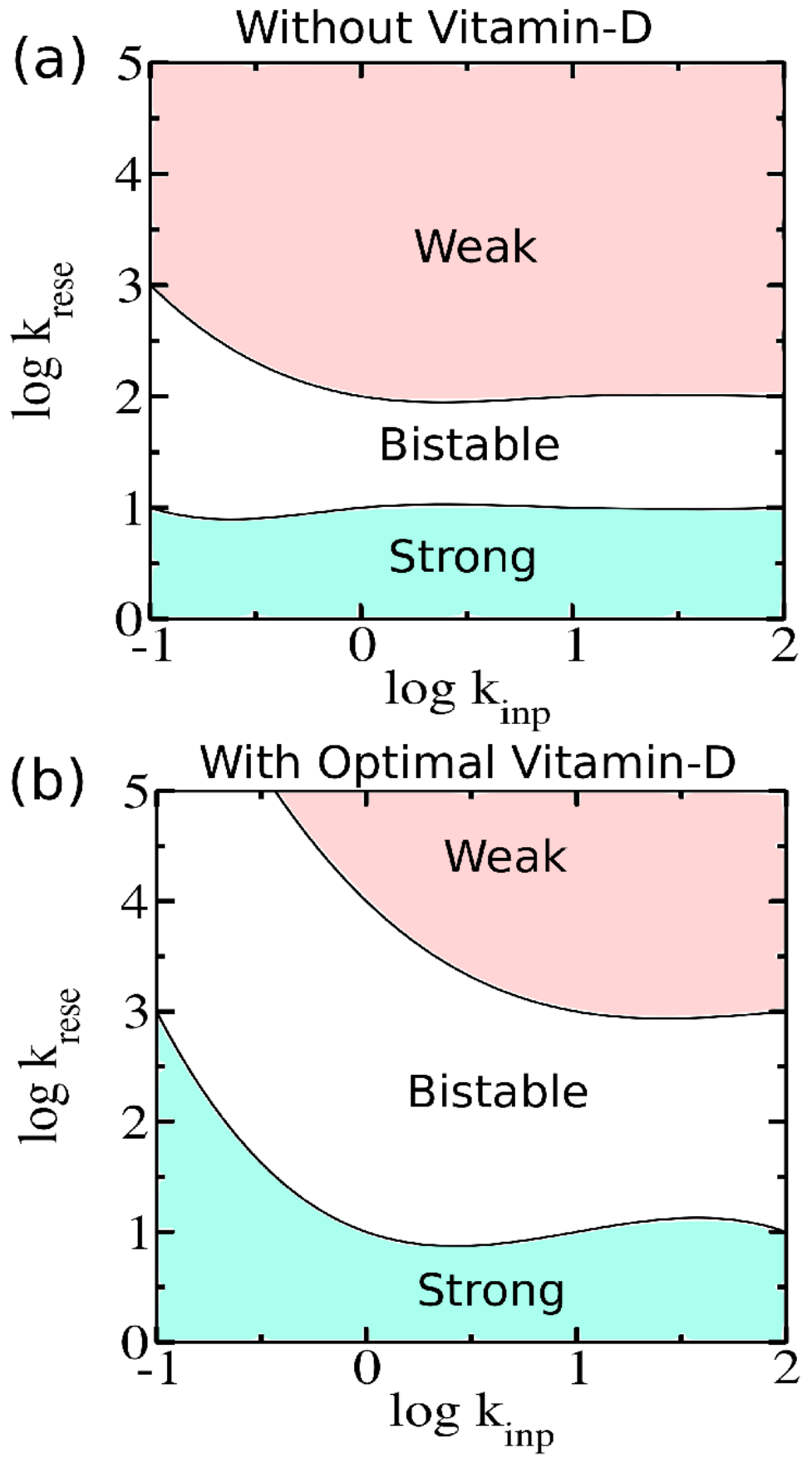

**Figure 3: Impact of vitamin-D over the phase diagram of immune regulation.**

**(a) In absence of vitamin-D the pair of regulation rates, (i) pathogenic stimulation $(k_{inp})$ and (ii) the resting APC mediated effector T-cell regulation $(k_{rese})$ are varied to find the boundaries between the two specific regulation regions: Weak and strong. These two regulation regimes are intervened by a bistable region where a moderate regulation is operated. (b) The same way in presence of vitamin-D the boundaries separated regions are alienated. Note that in presence of vitamin-D boundaries are shifted: strong regions become broader. Bistable regions become relatively expanded. Weak regions become considerably compressed than what happens in absence of vitamin-D. The rate constants considered here are similar to Figure 2.**

## Time evolution of immunological components

### In absence of (or at very low concentration) of Vitamin-D

We observed some interesting results from the study of the time evolution analysis of the immunological components in the above mentioned three regulation regions. Here we have presented the dynamical changes of elements against time (days) that quantitatively explain some attributes of the immune responses both in absence (see **Figure 4(a)**) and presence of vitamin-D (see **Figure 4(b)**). In **Figure 4(a)** within few hours we see that there is a sudden increase in the amount of effector T-cells which reaches to a peak value. This, as said before, is typically referred to the onset of an adaptive response. This is in common agreement with most experimental results which suggests that recognition takes place within few hours after the pathogenic incursion [42, 43].

We see in **Figure 4** that the pathogenic growth starts dying out at a much faster rate immediately after the initiation of effector T-cell production. We now have a huge population of effector T-cells that have been activated from the naive T-cells after APC activation. The population of these T-cells remains considerably higher even after the population of pathogen becomes significantly less. Such explosion in effector T-cell production thus often causes various kinds of autoimmune diseases [13-15, 44].

After the death of incoming pathogen *a new steady state* is developed at longer time. Also, once steady state has been reached after 10-20 days or so, one can notice the steady state values of effector T cells. This is in accordance with the fact that these T-cells are antigen specific cells. Once the pathogen is suppressed, the body creates an immunological memory of that specific pathogen, which corresponds to a steady state value of effector T cells. It might particularly be useful if the same pathogen strikes again. Then the amount of antigen specific T-cells grows fast which would lead to a rapid and more effective suppression of targeted pathogens [45].

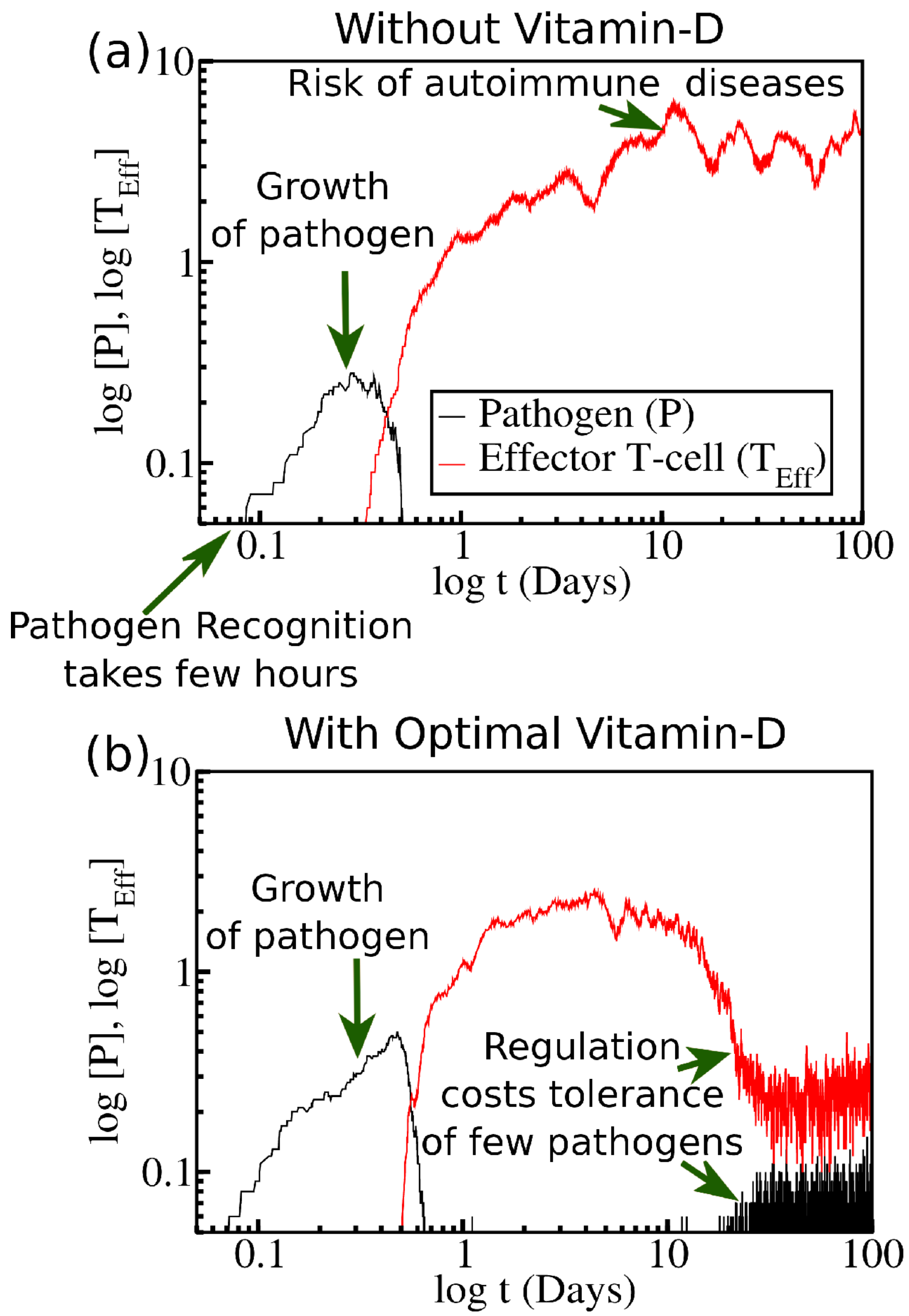

**Figure 4: Time evolution of immune response.**

**The dynamical variation of pathogens and efffector T-cells are calculated both (a) in absence and (b) in presence of optimal level of vitamin-D. In both cases adaptive response sets in after few hours of the pathogenic incursion. Pathogens are destroyed when effector T-cells reaches its peak value. Although regulatory T-cells control a little over the huge production of effector T-cells even after the death of pathogen and its large steady state concentration may increase the risk of autoimmune diseases. The presence of active vitamin-D subsequently controls over the large**

**production of effector T-cells. The steady state concentration of effector T-cells now low down significantly which may decrease the risk of autoimmune diseases. At the same time, note the re-entrant possibility of pathogen which up to certain level assists to build an adaptive tolerance of the immune system. (In both cases** $k_{inp}, k_{rese}$ **are so chosen that they remain in the bistable region** $(k_{inp} = 10, k_{rese} = 100)$ **as shown in Figure 3. The other rate constants considered here are similar to Figure 2.**

### In the presence of standard level of Vitamin-D

Vitamin-D plays a crucial role on the onset of adaptive response. It modifies the scenario as we have explained in the last subsection. Once the T-effector population starts increasing, production of active vitamin D [D*] is upregulated. This, in turn, upregulates regulatory T-cell growth, which along with [D*] regulate the aggressive, inflammatory responses exerted by T-effector cells, restoring order to the body. However, in this process, T-cell population relaxes at a much faster rate (see **Figure 4(b)**). As a result, rate of pathogen killing is significantly suppressed. In our model vitamin-D activation starts to grow rapidly within day 1 or 2. Hence we find that active vitamin-D does not play any substantial role in the very initial stage of pathogenic growth or decay. In presence of vitamin-D we observe a re-entrant possibility of pathogen which may sustain for long time [12].

In addition, we have studied pathogen relaxation dynamics at different vitamin-D concentrations. We have observed that if there would not be any vitamin-D present in our body, pathogenic growth would have been more effectively destroyed by the explosive production of effector T-cells immediately after recognition. But to compensate the large population of effector T-cells, vitamin-D plays a role as negative catalyst in effector T-cells production. As a consequence, in presence of vitamin-D pathogen annihilation rate at longer time also become suppressed. Hence, we find from the above analysis that in order to stay away from autoimmune diseases while we switch on the action of optimal regulation by active vitamin-D, at the same time our immune system becomes bound to tolerate some amount of pathogen. In fact, a healthy immune system is always characterized by tolerating certain extent of pathogenic stimulation. The fact, that vitamin-D has been implicated as an important factor in several different autoimmune diseases by preserving bistability, suggests its efficiency in controlling body's self-tolerance [46-48]. It is worth mentioning here that experimental observations related to the adoptive transfer of tolerance also supports the emergence of such bi-stability where the balanced co-existence of strong and weakly regulated immune responses is preserved in the system [49]. However in presence of very high concentration of vitamin-D or any of its steroid analogues which exert such an over-strong regulation that pathogenic resistance of immune system is totally failed. As a consequence, it induces the risk of pathogenic re-attack.

**Steady state analysis and optimal vitamin-D**

One important detail that needs to be considered here is the emergence of the new steady state in presence of vitamin-D with its tightly controlled homeostasis. To understand the relevance of vitamin-D in the above response, different initial concentrations of vitamin-D, [$D_{in}^0$] were considered. We have thus considered the initial concentrations of vitamin-D from lower to higher than the steady state value of vitamin-D, from $10^{-4}$ to $10^4$ nmol/lit. The variation of T-cell levels and pathogen levels in the newly established steady state were obtained and these concentrations are plotted versus log [$D_{in}^0$] in **Figure 5**. To measure a safe vitamin-D range we need to control the immune-regulation as well as pathogenic resistance as these are intimately connected. It is important to note that we cannot establish such a strong regulation by vitamin-D beyond which pathogenic resistance by effector T-cell subtly fails.

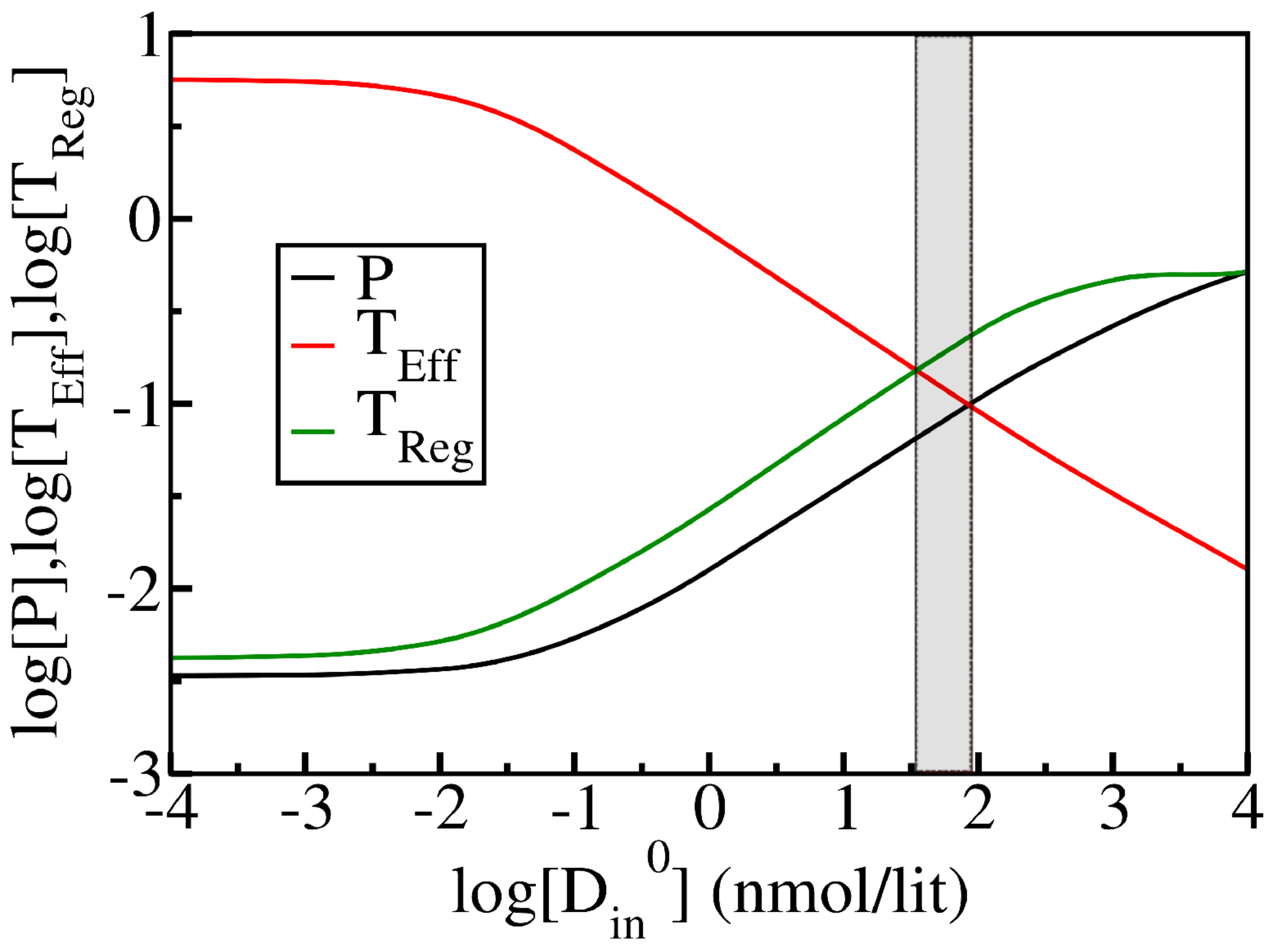

**Figure 5: Steady state value analyses as a function of log (initial vitamin-D level).**

**(a) Evaluation of steady state regulation in terms of effector T-cell and regulatory T-cell concentration at different initial intake of vitamin-D. At certain range of vitamin-D concentration (around the value of vitamin-D concentration above 50 nmol/lit) $T_{Eff}$ concentration falls than the value of $T_{Reg}$ to establish a strong regulation that is necessary for prevention of autoimmune diseases. (b) Steady state evaluation of pathogenic resistance in terms of both pathogen and effector T-cell concentration at different initial intake of vitamin-D. At vitamin-D concentration beyond 50 nmol/lit when system gears up a more strong regulation, the steady state value of pathogen starts on increasing even exceeding the value of $T_{Eff}$ beyond $[D_{in}^{0}]$ =100nmol/lit. However we indicate (with a gray limit bar) the optimal vitamin-D ranges from 50 to 100nmol/lit where both pathogen and effector T-cell level remain at reasonably low value. Vitamin-D level beyond 100nmol/lit corresponds to an alarming concentration compared to the standard vitamin-D limit.**

The effects of local conversion of inactive $25(OH)D_3$ to active $1,25(OH)_2D_3$ by DCs on subsequent T cell responses were measured by flow cytometry and the results were extensively analyzed by Jeffery et al. [50]. They studied how this conversion can promote an anti-inflammatory T cell phenotype (such as CTLA-4) and inhibit the inflammatory expression of IL-17, IFN-γ, and IL-21. The dose dependent variations of such T-cell responses were shown in Fig. 2.(F) in the referred article [50]. The trend of responsive changes along with the concentration of $25(OH)D_3$ matches fairly well with the results depicted in **Figure 5** that we obtain from our model calculation. Following their cue, in the present study we also consider the circulating inactive form of vitamin D ($25(OH)D_3$) as an efficient marker of vitamin D status. Our dose dependent curves also match with the experimental findings of Correale et al. [25].

For the above data set, we find that the optimal vitamin-D level lies in the 50-100 nmol/lit range where both pathogen and effector T-cell levels remain at reasonably low risk range. Recently a large number of epidemiological studies and an U.S. Institute of medicine committee reported that a serum 25-hydroxyvitamin-D level of >20 ng/mL (50 nmol/L) is desirable for bone and overall health [51-53]. Those studies recommend both the upper and the lower limits of safe vitamin-D intake. High IgE levels were seen at very low 25-hydroxyvitamin-$D_3$ (<10 ng/mL or, <25 nmol/L) and at very high 25-hydroxyvitamin-$D_3$ (>135 nmol/L) levels [52].

Another important study found that high $25(OH)D_3$ concentration (greater or =100 nmol/L) often leads a statistically significant (2-fold) enhancement of pancreatic cancer risk [53, 54].

Therefore, the present study provides an estimate in the right range of optimal vitamin-D concentration.

**Sensitivity towards vitamin-D associated parameter set**

To investigate both the robustness and the sensitivity of vitamin-D related rate constants, it is essential to scrutinize their effects in a wide ranging scale. An additional reason to substantiate the sensitivity is that these values vary from system to system (here person to person) and the values can fluctuate even for the same person depending on various conditions. Though precise number of the rate constants may vary, the effective trend ought to preserve within a certain scale.

As both the active APC and effector T-cells are modulated by the impact of active vitamin-D we have investigated the outcome of different possibilities of the combination of $k_{aD*}$ and $k_{eD*}$ (defined in **Table 1**). From **Figure 5** we learned that to obtain a safe boundary of vitamin-D impact we need to efficiently check both effector T-cell concentration as well as pathogen concentration. Here we have scanned the parameter space to distinguish different zones observing the population of pathogen and effector T-cells. However at high vitamin-D concentration, pathogen may enhance their growth. Here the parameter space $\left(\log k_{aD*}, \log k_{eD*}\right)$ suggests that pathogenic and effector T-cell profile is less sensitive towards $k_{eD*}$. It rather shows a significant variation with the change of $k_{aD*}$. This analysis shows two distinct regions: (i) In the region of low vitamin-D impact $\left(k_{aD*} : 10^{-8} - 10^{-4}\right)$ we obtain *pathogen defeated zone* where pathogen concentration is found to be negligible but at the same time here in the range of $\left(k_{aD*} : 10^{-8} - 10^{-7}\right)$ we found a effector *T-cell flare-up zone*. (ii) In the region of very high vitamin-D impact $\left(k_{aD*} : 10^{-1} - 10^{2}\right)$ we obtain *effector T-cell defeated zone*. Here we find a *pathogen relapsing zone* where the steady state concentration of pathogen remains significantly large when the system is hyper regulated. It'll be safe if we stay in between $k_{aD*} : 10^{-7} - 10^{-4}$ and also $k_{eD*} : 10^{-6} - 10^{-2}$ to avoid high pathogenic and effector T-cell population (see **Figure 6**).

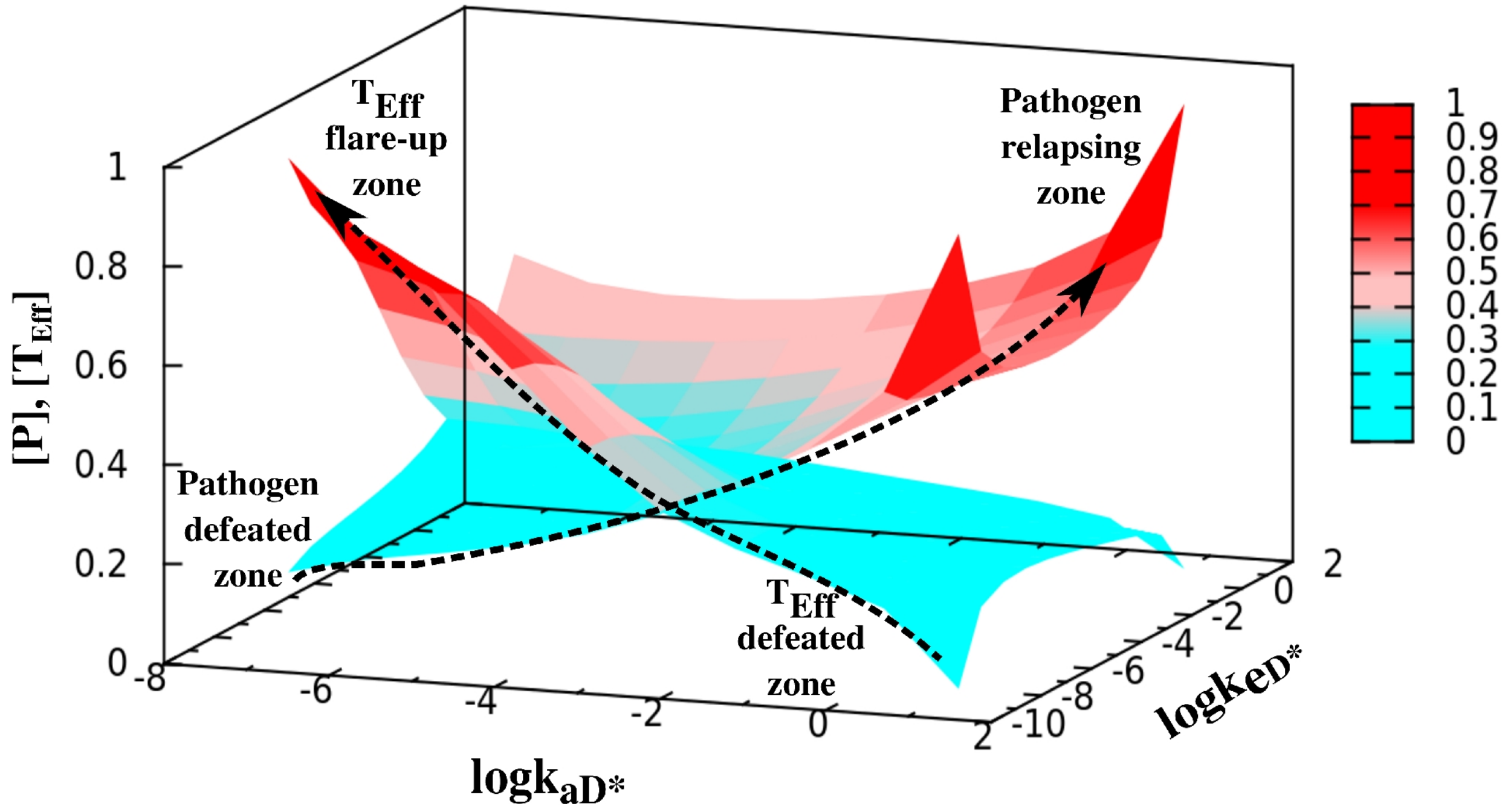

**Figure 6: Impact of vitamin-D over the steady state profiles of pathogen, [P] and effector T-cell, [$T_{Eff}$].**

**(a) We vary simultaneously the impact of [D*] over APCs $\left(k_{aD*}\right)$ and effector T-cells$\left(k_{eD*}\right)$. We find different regions: (i) At low vitamin-D impact $\left(k_{aD*} : 10^{-8} - 10^{-4}\right)$ we obtain pathogen defeated zone but $T_{Eff}$ flare-up zone$\left(k_{aD*} : 10^{-8} - 10^{-7}\right)$. (ii) At high vitamin-D impact$\left(k_{aD*} : 10^{-1} - 10^{2}\right)$ steady state concentration of pathogen largely increases which distinguished as pathogen relapsing zone. In pathogen relapsing zone, however we find $T_{Eff}$ defeated zone. The basic value parameters are taken as $k_{inp} = 10, k_{rese} = 10^{2}$. Other parameter values are taken from Table 1.**

## Discussion and Summary of Results

As already mentioned recent experiments provided quantitative information on the immuomodulatory functions of vitamin-D and established functions of this vitamin beyond its well-stated role in calcium metabolism [19-25]. To understand these recent experiments, we developed a theoretical coarse-grained model based on interaction network. The network dynamically connects different immune components that are experimentally found to be involved in the vitamin-D regulated immune responses. The formulated kinetic scheme describes the time evolution of these components that mainly include pathogen, vitamin-D, APCs, effector T cells, regulatory T cells. Here we summarize the pertinent observations that emerged from the kinetic network model.

(i) The steady state analysis of the present kinetic scheme finds the three regulation limits: weak, moderate and strong, both in absence and presence of vitamin-D. The phase diagrams of boundary separated three immune regulation regions show that in presence of optimal vitamin-D, strong regulatory region becomes broad and in the moderate (or, bistable) regulatory region becomes more extended. The weak regulatory region shifts towards the large values of APC mediated effector T-cell activation rate ($k_{rese}$) and becomes more constricted than what happens in absence of vitamin-D. This investigation offers a semi-quantitative picture supporting several experimental and clinical observations that show how vitamin-D prevents the immune system from any sort of autoimmune diseases by restricting its function within strong to moderate regulation limits.

(ii) The analysis of time evolution of immunological components explicitly shows the attainment of a new steady state in the presence of optimal level of vitamin-D. The dynamical characterization of the involved components reveals that the recognition of the pathogenic growth requires a few hours and this fact is in general agreement with most experimental results. After the activation of vitamin-D, the population blast of effector T-cells relaxes to a comparatively lower value (as and when we include the effects of optimal vitamin-D). But such downward regulation for the prevention of autoimmune diseases costs re-entrant possibilities, to certain extent, of pathogen which again enhances the tolerance capability of a healthy immune system. The importance of vitamin-D in control of tolerance has also been experimentally verified.

(iii) Quantitative predictions of the present model are in good agreement with several recent experimental studies and clinical observations [12, 25, 42-48, 50-56]. We have attempted to quantify how much vitamin-D is needed to resist autoimmunity and why. Our dose dependent variations of T-cell responses along with the concentration of vitamin-D seem to have an excellent correlation with experimental findings of Jeffery et al. and Correale et al. [50, 25]. *We additionally find that a safe range of vitamin-D is essentially determined by the interrelated level* *of pathogen, effector T-cells and regulatory T-cells.* The range is restricted by both hyper-regulation and effector T-cell inflammation. Very recent randomized controlled trials (RCTs) suggest that there should be an element of caution about recommending high serum 25(OH)$D_3$ concentrations as routine clinical practice and that should spread among the entire population. Thus both experimental and theoretical initiatives should begin to emerge by the above fact [54-56].

(iv) The regulatory impact of active vitamin-D over APC and effector T-cells is investigated here by steady state analysis. We find that the nonlinear regulation of vitamin-D is sensitive towards APC functioning. This particular impact parameter largely controls the emergence and the range of bistability. Early experimental studies also report such markedly affected DC maturation and activation profile in presence of vitamin-D [21].

As we mentioned before, the steady state analysis of the proposed master equations reveals intricate relations between vitamin-D levels and T-regulatory cells maintained by homeostasis. These relations suggest that at homeostasis, lower levels of [$D_{in}^0$] correspond to a lower population of T-regulatory cells, which again suggests that once a pathogen/antigen enters the body, the nature of the immune response is expected to be less regulatory and hence more inflammatory or aggressive. In addition in a weak regulation limit we have studied the temporal progression of both regulatory and effector T-cells. Interestingly, we find coupled oscillatory dynamics of effector T-cells ($T_{Eff}$) and regulatory T-cell ($T_{Reg}$) that begin to develop within 2-5 days and periodically continue. In the presence of pathogen when the system tends towards a slightly weak regulation regime we observe a dynamic cross regulation in the temporal progression of regulatory and effector T-cells population. This is presented in the supplementary material (see **Figure S1**). The impact of vitamin-D associated intrinsic oscillatory behavior over effector T-cells could provide a dramatic signature of disease phenotype in clinical therapy [29].

The critical role of the various cells involved in immune response, especially [$D_{in}^0$] and [D*] concentration could be understood via investigating dynamics of response. We are indeed aware of the fact that quantitative results of in-vivo analysis of the effects of the high dose vitamin D level or its any steroid analogue are somewhat ambiguous. The consequences of both low and *very* high dose of vitamin D causing fatal diseases are relatively well established. We are also aware of the persisting current dilemma of precisely defining the vitamin D insufficiency and difficulty in identifying the safe range. Our model calculation efficiently quantifies that there exist a delicate window of concentrations of vitamin-D which would be critical in maintaining an appropriate response to a pathogen. Extremely low levels of vitamin-D could lead to increased risk of autoimmune responses and extremely high levels would suggest an extremely tolerant response, which could increase the risk of tumors and cancerous cell growth and various allergic responses stimulated by the elevated IgE concentrations [55, 56].

It is important to note that two enzymes CYP27B1 and CYP24A1 and the population of VDR play important role in balancing several immunological responses. Defect in or unavailability of any of these proteins will greatly perturb the whole immunological network. A series of D*-VDR mediated processes that have enormous consequences have not been fully understood yet. Malfunction of these enzymes (such as: CYP27B1 and CYP24A1) can also reflect a deeper problem (such a genetic) that is difficult to rectify [57, 58]. It clearly needs a more quantitative analyses.

It is worth mentioning here that the activation of a naive T cell into an effector or regulatory T cell is also a complex process. This begins with the scanning of the surface of the APC's in the lymph nodes for the MHC class II type molecules by the naive T cells. If a particular epitope is recognized and co-stimulatory molecules are present, then the activation process is initiated [59, 60]. Now this can be understood via an energy landscape analysis. The process of successful activation can be thought of as the T cell negotiating a barrier in the energy landscape. This can be brought about through either a single successful contact with an APC or multiple contacts if the second or later contact occurs within a finite time. If the T-cell is above the seperatrix in the energy landscape then the probability of a successful activation is higher which is only present for a finite time after the previous excitation. The above picture is similar to the immunological studies carried out by Hong et al. [61] and Das et al. [60] and the enzyme catalysis model proposed by Min, Xie and Bagchi earlier [62]. However, to make the present model tractable, we had to ignore such complexity of T-cell activation.

The master equation approach adopted here can be and has been solved both by a deterministic and a stochastic approach, given the initial values of the parameters and the fluxes. Within a biological cell, there can always be large fluctuations due to environmental factors or other causes [63, 64]. Such fluctuations can induce the cross-over from weak regulation to strong regulation. This is an issue that deserves further study.

Although our model is coarse-grained and the evaluated results are semi-quantitative due to absence of some kinetic parameters, this study, perhaps, constitutes the first theoretical investigation of the role of vitamin-D in immune regulation. Despite its limitations, we believe that the kinetic interplay between pathogen, effector T-cells and the unavoidable participation of vitamin-D to remain the basic ingredients in the upcoming studies.

In future, we plan to extend our system of equations to include effects of drugs such as immune suppressants such as glucocorticoids that introduce a further competition in the reaction network [13]. The role of immune suppressants is not fully understood but known to exhibit serious side effects like cancer.

# Methods

### Coarse-grained reaction network model development

As already mentioned, in order to describe the complex interplay among different types of immune cells, pathogens and the modulatory role of vitamin-D, we need to develop in the beginning a simple coarse-grained approach that can both be solved and understood. The complexity arises because of the large number of biochemical machineries in the human body that are strongly coupled with each other [65, 66]. Understanding the relationship between these different machineries involving different types of cell may ultimately require detailing at the molecular level. A simpler, albeit cruder version is proposed here that accounts for some of the complexities present at the molecular levels by coarse-graining them at the cellular level. A pictorial description of initial complex network and the associated coarse-grained network are demonstrated in the supplementary material (see **Figure S2 and S3**). With this goal in mind, we perform model analysis based on T-cell activation, deactivation and regulation, following some experimental results discussed below.

(i) Myeloid dendritic cells (here we call them as antigen presenting cell (APC)) which are present in different organs are the key players involved in triggering the onset of an adaptive response. While pathogen phagocytosed dendritic cells serve to activate naive T cells into effector T cells, immature dendritic cells upon pathogen contact convert naive T cells into regulatory T cells [19, 20].

(ii) Effector T-cells (described in the supplementary material, S1) release cytokines which upregulate the activity of 1α-hydroxylase enzyme (CYP27B1), which in turn increases production of active vitamin-D [D*] from its inactive form. It is worth mentioning here the extensive experimental study by Correale et al. which reported that CD4+ T cells are capable of metabolizing 25(OH) Vitamin D to 1,25 $(OH)_2$ Vitamin D, which again inhibits T cell function [25].

(iii) Active form of vitamin-D, 1, 25$(OH)_2D_3$ modulates the immune response through the inhibition of DC differentiation and maturation into potent APC [21].

(iv) While effector T cells are directly inhibited by the increased production of [D*], CD4+CD25+ FoxP3+ regulatory T cells ($T_{Reg}$) cells become up-regulated. In addition, these $T_{Reg}$ cells also efficiently inhibit T-effector cells proliferation [22-24]. However, previous experimental data confirms that IL-10 is a positive autocrine factor. They act directly on effector T cells even in the absence of antigen presenting cells, enhancing the regulatory action of 1,25 $(OH)_2$Vitamin D [25].

Coarse-graining of the interaction network is accomplished through making a few simplifying observations and vital assumptions. They are as follows:

(a) Th1, Th2 and Th17 cells are grouped together as effector T-cells. The detailed description of these T-cells is depicted in the supplementary material (S1) [16].

(b) It is well established that the primary molecular action of 1,25$(OH)_2D_3$ is to initiate gene transcription by binding to VDR which is a member of the steroid hormone receptor superfamily of ligand-activated transcription factors. VDR therefore is an important factor in 1,25$(OH)_2$ $D_3$ mediated functions. More detailed information about the molecular activation of VDR and its role in gene transcription can be obtained from the recent review appraised by Pike et al. [67].

On contrary, we found that there are growing evidences that 1,25$(OH)_2D_3$ also has rapid actions that are not essentially mediated through transcriptional events involving VDR. They are in fact membrane initiated actions [68]. In the present model we have not included the effect of VDR. We have only considered the production of active vitamin-D from its inactive form upon T-cell activation.

(c) In circulation, the inactive form of vitamin-D, 25$(OH)D_3$, is generally used as an indication of vitamin-D status. However, in dendritic cells (DC) use of this precursor depends on its uptake by cells and subsequent conversion by the enzyme CYP27B1 into active 1,25$(OH)_2D_3$ [69]. It is represented as [D*]. Active Vitamin-D, [D*], has a tight control over the homeostatic production rate that auto-regulates its production by directly upregulating the activity of the P450 cytochrome CY24A1. In our model we have considered the steady state rate of inactive vitamin-D that found from experimental and clinical measurements while keeping the concentration of these enzymes as the implicit factors.

In the present context we consider the following set of biological transformations. Most of them are catalytic reaction in terms of up-regulation or down-regulation.

(1) The primary step is the annihilation of pathogen by effector T-cell.

$$\text{Pathogen }\left(\text{P}\right)+\text{Effector T-cell }\left(\text{T}_{\text{Eff}}\right)\rightarrow\text{P}_{\text{killed}}+\text{T}_{\text{Eff}} \qquad \text{(i)}$$

(2) Production of effector T-cell requires the presence of active antigen presenting cell (APC). Active APC, on the other hand is produced by following sequence of reactions. $1^{st}$ resting APC forms through the interaction between inactive APC and pathogen.

$$\text{Inactive APC }\left(\text{A}_{\text{in}}\right)+\text{P}\rightarrow\text{Resting APC }\left(\text{A}_{\text{Res}}\right)+\text{P} \qquad \text{(ii)}$$

(3) Further pathogenic contact and/or effector T-cell contact promotes the resting APC to turn out to be active APC.

$$\left\{\begin{array}{l}\text{A}_{\text{Res}}+\text{P}\rightarrow\text{A}_{\text{Act}}+\text{P}\\ \text{A}_{\text{Res}}+\text{T}_{\text{Eff}}\rightarrow\text{A}_{\text{Act}}+\text{T}_{\text{Eff}}\end{array}\right\} \qquad \text{(iii)}$$

(4) Then effector T-cell is produced by the interaction between precursor/native T-cell with active APC.

$$\text{Native T-cell }\left(\text{T}_{\text{Nat}}\right)+\text{Active APC }\left(\text{A}_{\text{Act}}\right)\rightarrow\text{T}_{\text{Eff}}+\text{A}_{\text{Act}} \qquad \text{(iv)}$$

(5) Simultaneously inactive vitamin-D is transformed into active vitamin-D upon effector T-cell contact.

$$\text{Inactive Vitamin-D }\left(\text{D}_{\text{in}}\right)+\text{T}_{\text{Eff}}\rightarrow\text{ActiveVitamin-D }\left(\text{D*}\right)+\text{T}_{\text{Eff}} \qquad \text{(v)}$$

(6) Resting T-cells and Vitamin-D, both can initiate the formation of regulatory T-cell from native T-cell.

$$\text{Native T-cell }\left(\text{T}_{\text{Nat}}\right)+\text{Resting APC }\left(\text{A}_{\text{Res}}\right)\rightarrow\text{T}_{\text{Reg}}+\text{A}_{\text{Res}} \qquad \text{(vi)}$$

$$\text{Native T-cell }\left(\text{T}_{\text{Nat}}\right)+\text{Active Viatmin-D }\left(\text{D*}\right)\rightarrow\text{T}_{\text{Reg}}+\text{D*}$$

(7) Both Regulatory T-cell and active vitamin-D annihilate effector T-cell to control the hyperactivity of the immune system.

$$\left\{\begin{array}{l}\text{T}_{\text{Eff}}+\text{T}_{\text{Reg}}\rightarrow\text{T}_{\text{Eff}}^{\text{killed}}+\text{T}_{\text{Reg}}\\ \text{T}_{\text{Eff}}+\text{D*}\rightarrow\text{T}_{\text{Eff}}^{\text{killed}}+\text{D*}\end{array}\right\} \qquad \text{(vii)}$$

(8) The cycle is completed by the transformation of active APC to resting again by the same duo, $T_{Reg}$ and D* which work at tandem.

$$\left\{\begin{array}{l} A_{Act} + T_{Reg} \rightarrow A_{Res} + T_{Reg} \\ A_{Act} + D^* \rightarrow A_{Res} + D^* \end{array}\right\} \quad \text{(viii)}$$

**Master equations quantifying the reaction network dynamics**

Now, some important further assumptions before we set about writing the master equations:

(i) For Pathogen/Antigen, inactive APC and native T-cells, each has a birth rate which includes influx and proliferation rates and a death rate similar to decay which incorporates natural cell death. The death rate of each component is linear with its concentration.

(ii) The transition probabilities are all assumed to be constant with time but may vary from system to system (i.e. here person to person) according to the condition applied.

(iii) To scale the unit, here we assume that in absence of pathogen, hundred (average number of T-cell present in hundred nano-liter blood sample) precursor T-cells pre-exist.

The above annihilation, recombination and catalytic reactions lead to the following set of coupled master equations. The equations are size-extensive. In fact the size extensibility is the critical robustness of our model.

$$\frac{dP}{dt} = \sigma_P - \pi_P P - k_P T_{Eff} P \quad (1)$$

$$\frac{dA_{in}}{dt} = \sigma_A - k_{inp} A_{in} P - m_a A_{in} \tag{2}$$

$$\frac{dA_{Res}}{dt} = k_{inp} A_{in} P + k_{ar} T_{\mathrm{Reg}} A_{Act} + k_{aD*} A_{Act} D* - k_{rese} A_{Res} T_{Eff} - k_{inp} A_{Res} P - m_a A_{Res} \tag{3}$$

$$\frac{dA_{Act}}{dt} = k_{rese} A_{Res} T_{Eff} + k_{inp} A_{Res} P - k_{ar} T_{\mathrm{Reg}} A_{Act} - k_{aD*} A_{Act} D* - m_a A_{Act} \tag{4}$$

$$\frac{dT_{Nat}}{dt} = \sigma_T - k_{an} A_{Act} T_{Nat} - k_{resn} A_{Res} T_{Nat} - k_{nD*} T_{Nat} D* - m_n T_{Nat} \tag{5}$$

$$\frac{dT_{Eff}}{dt} = k_{an} A_{Act} T_{Nat} - k_{er} T_{Eff} T_{\mathrm{Reg}} - k_{eD*} T_{Eff} D* - m_e T_{Eff} \tag{6}$$

$$\frac{dT_{\mathrm{Reg}}}{dt} = k_{resn} A_{Res} T_{Nat} + k_{nD*} T_{Nat} D* - m_r T_{\mathrm{Reg}} \tag{7}$$

$$\frac{dD_{in}}{dt} = \sigma_D - k_{eD} T_{Eff} D_{in} - m_D D_{in} \tag{8}$$

$$\frac{dD*}{dt} = k_{eD} T_{Eff} D_{in} - m_{D*} D* \tag{9}$$

Where the terms signify as follows:

$k_{ij}$ → Transition probability rates,

$\sigma_K$ → Production rate by body of component $k$ ,

$m_i$ → Overall death rate of component $i$ ,

$P$ → Concentration of Pathogen/Antigen/Self-Antigen etc,

$A_{in}$ → Concentration of inactive antigen presenting cells without pathogen capture,

$A_{Res}$ → Concentration of resting antigen presenting cells after pathogen capture,

$A_{Act}$ → Concentration of activated antigen presenting cells after pathogen recognition and effector T-cell contact.

$T_{Nat}$ → Concentration of naive antigen-specific T cells,

$T_{Eff}$ → Concentration of effector T cells,

$T_{\mathrm{Reg}}$ → Concentration of regulatory T cells,

$D_{in}$ → Concentration of Inactive form of Vitamin-D3 (**1, 25(OH)D**) in the body

$D^*$ → Concentration of active form of Vitamin-D3 (**1, 25$(OH)_2$D**) in the body

That is, we have used the same letter to denote both the species and its concentration. This should not cause any confusion.

## System parameters and data analysis

A set of nine coupled differential equations is difficult to solve analytically. We obtain the time dependent concentrations of all the components involved in the scheme by employing the well-known stochastic simulation analysis proposed by Gillespie [38]. Both the single molecular as well as ensemble enzyme catalysis have been studied following this method. All the results presented in this article are derived using stochastic simulation method. However, we have also verified the consistency of each result by using the deterministic approach which is easier to implement.

Here we have considered one hundred nano-litre volume of blood sample. In the absence of pathogen this blood sample effectively contains the steady state concentration of all the precursor cells. [70-72] Since all the reactions are bimolecular, the volume dependence of the reaction is expected to be an issue. Thus, we have kept fixed the box volume to one hundred nano-liter and all the rate constants are in the unit of per day. We have closely followed the type of formalism developed in Ref. 30.

Furthermore, we have assumed that in the absence of antigen, hundred precursor T-cells can pre-exist within this fixed volume (100 nano-litre), in accord with known experimental values [70, 71]. These T-cells have a 1% turnover per day. Concentrations of pathogens and APCs are also normalized. The production rate and death rate of these components are so assigned that their steady state values become one. Other associated probabilities/rate constants of different reaction sets are used from early papers in this field [30]. However, for vitamin D, the production and mortality rate constants are calculated from their steady state concentration. Other vitamin D related rate constants are treated as variable in our study, as we have no experimental data available on them. In reality, such model requires to estimate several rate parameters values. Accurate values of some of these rate constants are unfortunately very hard to determine. Such rate parameters depend on several factors and differ from species to species. So they do not have any specific standard value. As for example, it would be quite difficult to determine the mortality rate of effector and regulatory T-cell as in the present model these rate parameters also include the proliferation rate along with their death rate. Moreover the pathogenic stimulation could be at various ranges according to their strength and pattern.

Hence the primary difficulty of predictive theoretical research in this area is the absence of accurate values of rate constants/transition probabilities. In the present study we have employed the following approach to circumvent this difficulty. (i) In some cases where values could be estimated from literature, we have used the known value and varied it over a range to check the sensitivity of results. (ii) In a few cases, order of magnitude estimates for values were employed. [30] We also focused on exploring the phase diagram by varying some key rate parameters that are not known and looked for the optimum region where results are sensitive to the parameter space (given experimental and assumed values of the rate constants and concentrations). To this end, we have varied the rate constants over a significantly wide range. In addition, the concentration of precursor elements was normalized, so as to reflect manifold change in the production level. Taking typical values as mentioned below (see **Table 1**), the time evaluation of the system and other analyses are performed in the present work. Here we have used the standard definition of steady state, i.e.; when the concentration of different species is invariant with time (dc/dt=0). In particular, for stochastic simulation, a steady state is assumed to reach when the concentration of a species fluctuates around a mean value without any noticeable drift at long time.

**Table 1: Basic parameter values (*time duration is taken as "days").**

| Parameter | Symbol | Value |
|---|---|---|
| **Reproduction rate of pathogen** | $\sigma_P$ | **1** |
| **Death rate of pathogen** | $\pi_P$ | **1** |
| **Birth rate of APC** | $\sigma_A$ | **0.2** |
| **Death rate of APC** | $m_a$ | **0.2** |
| **Rate of pathogen killing by efffector-T cells** | $k_P$ | **100** |
| **Rate of APC activation by pathogen** | $k_{inp}$ | **Variable** |
| **Rate of APC reactivation by effector T cells** | $k_{rese}$ | **Variable** |
| **Rate of APC inhibition by regulatory T cells** | $k_{ar}$ | **$10^{-1}$** |
| **Rate of APC inhibition by active vitamin-D** | $k_{aD*}$ | **Variable** |
| **Birth rate of Native T cells** | $\sigma_T$ | **1** |
| **Rate of differentiation of native T cell to effector T cell induced by active APC** | $k_{an}$ | **1** |
| **Rate of differentiation of native T cell to regulatory T cell induced by resting APC** | $k_{resn}$ | **1** |
| **Mortality rate of native T cell** | $m_n$ | **0.01** |
| **Rate of inhibition of effector T cell by active vitamin-D** | $k_{eD*}$ | **Variable** |
| **Rate of inhibition of effector T cell by regulatory T cell** | $k_{er}$ | **10** |
| **Rate of decay of effector T cells** | $m_e$ | **0.1** |
| **Rate of regulatory T cell reactivation by active vitamin-D** | $k_{nD*}$ | **$10^{-7}$** |
| **Rate of decay of regulatory T cells** | $m_r$ | **0.1** |
| **Production rate of inactive vitamin-D** | $\sigma_D$ | **1** |
| **Death rate of inactive vitamin-D** | $m_D$ | **$10^{-9}$** |
| **Rate of reactivation of active vitamin-D induced by effector T cells** | $k_{eD}$ | **$10^{-7}$** |
| **Rate of deactivation of active vitamin-D** | $m_{D*}$ | **$10^{-2}$** |

## Acknowledgments

It is a pleasure to thank Prof. Anjali A. Karande, Dr. Mantu Santra, Dr. Biman Jana, Ms. Gauri Ranadive, Mr. Manoj Kumar Mahala and Mr. Kushal Bagchi for many helpful discussions. This work was supported in parts by grants from DST (India) and Sir J.C. Bose Fellowship (DST).

**Authors Contribution**

S.R. and B.B. designed research; S.R. and K.S. performed research; S.R. and B.B. analyzed data; and S.R. and B.B wrote the article.

**Conflict of Interest**

Authors don't have a conflict of interest.